# Toward inertial confinement fusion energy based on heavy ion beam

S. Kawata[1, *], M. Okamura[2], K. Horioka[3], C. Deutsch[4], T. Karino[5], D. H. H. Hoffmann[6, 7], Y. Zhao[6], S. Ikeda[2], T. Kanesue[2] and A. I. Ogoyski[8]

[1]Graduate School of Eng., Utsunomiya Univ., Utsunomiya 321-8585, Japan
[2] Collider-Accelerator Dpt., Brookhaven National Lab., Upton, NY 11973, USA
[3]Dpt. Energy Sciences, Tokyo Inst. of Tech., Yokohama 226-8502, Japan
[4]Laboratoire de Physique des Gaz et des Plasmas (LPGP), Université Paris-Saclay, 91405 Orsay, France
[5]Collaborative Laboratories for Advanced Decommissioning Science, Japan Atomic Energy Agency, Fukushima 970-8026, Japan
[6] Xi'An Jiaotong University, Xi'an, Shaanxi, 710049, P. R. China
[7]Inst. für Kernphysik, Technische Univ. Darmstadt, 64289 Darmstadt, Germany
[8]Dpt. of Physics, Technical Univ. of Varna, Studentska 1, Varna, Bulgaria

*Corresponding author: kwt@cc.utsunomiya-u.ac.jp

## Abstract

Heavy ion inertial fusion (HIF) energy would be one of promising energy resources securing our future energy in order to sustain our human life for centuries and beyond. The heavy ion beam (HIB) has remarkable preferable features to release the fusion energy in inertial confinement fusion: in particle accelerators HIBs are generated with a high driver efficiency of ~ 30-40%, and the HIB ions deposit their energy inside of materials. Therefore, a requirement for the fusion target energy gain is relatively low, that would be ~50-70 to operate a HIF fusion reactor with the standard energy output of 1GW of electricity. The HIF reactor operation frequency would be ~10~15 Hz or so. Several-MJ HIBs illuminate a fusion fuel target, and the fuel target is imploded to about a thousand times of the solid density. Then the DT fuel is ignited and burned. The HIB ion deposition range would be ~0.5-1 mm or so depending on the material. Therefore, a relatively large density-scale length appears in the fuel target material. The large density-gradient-scale length helps to reduce the Rayleigh-Taylor (R-T) growth rate. The key merits in HIF physics are presented in the article toward our bright future energy resource.

## 1. Introduction: Origin of HIF and Merits of HIB (Heavy Ion Beam) in ICF

The heavy ion beam (HIB) fusion (HIF) has been proposed in 1970's. Before HIF, a proton-beam inertial fusion target was proposed in Ref. [1]. The HIF reactor designs were also proposed in Refs. [2-5]. HIB ions deposit their energy inside of materials, and the interaction of the HIB ions with the materials are well understood [6, 7]. The HIB ion interaction with a material is explained and defined well by the classical Coulomb collision and a plasma wave excitation in the material plasma. The HIB ions do not reflect from the target material, and deposit all the HIB ion energy inside of the material. The HIB energy deposition length is typically the order of ~0.5-1mm in an HIF fuel target depending on the HIB ion energy and the material. When several MJ of the HIB energy is deposited in the material in an inertial confinement fusion (ICF) fuel target, the temperature of the energy deposition layer plasma becomes about 300 eV or so. The peak temperature or the peak plasma pressure appears near the HIB ion stopping area by the Bragg peak effect, which comes from the nature of the Coulomb collision. The total stopping range would be normally wide and the order of ~0.5-1 mm inside of the material. An indirect drive target was also proposed in Ref. [8]. However, due to the relatively moderate temperature inside of the material and the plasma temperature and pressure profiles, a spherical direct drive target would be appropriate in HIF.

In ICF, a driver efficiency and its repetitive operation with several Hz ~ 20 Hz or so are essentially important to constitute an ICF reactor system. HIB driver accelerators have a high driver energy efficiency of 30-40 % from the electricity to the HIB energy. In general, high-energy accelerators have been operated repetitively daily. The high driver efficiency relaxes the requirement for the fuel target gain. In HIF the target gain of 50~70 allows us to construct HIF fusion reactor systems, and 1MkW of the electricity output would be realized with the repetition rate of 10~15 Hz.

The HIB accelerator has a high controllability to define the ion energy, the HIB pulse shape, the HIB pulse length and the HIB number density or current as well as the beam axis. The HIB axis could be also controlled or oscillated with a high frequency [9-11]. The controlled wobbling motion of the HIB axis is one of remarkable preferable points in HIF, and would contribute to smooth the HIBs illumination non-uniformity on a DT fuel target and to mitigate the Rayleigh-Taylor (R-T) instability growth in the HIF fuel target implosion [12-15].

The relatively large density gradient scale length is created in the HIBs energy deposition region in an DT fuel target, and it also contribute to reduce the R-T instability growth rate especially for shorter wavelength modes [12, 15-17]. So in the HIF target implosion the longer-wavelength modes should be focused for the target implosion uniformity.

In ICF target implosion, the requirement for the implosion uniformity is very stringent, and the implosion non-uniformity must be less than a few % [18, 19]. Therefore, it is essentially important to improve the fuel target implosion uniformity. In general, the target implosion non-uniformity is introduced by a driver beams' illumination non-uniformity, an imperfect target sphericity, a non-uniform target density, a target alignment error in a fusion reactor, et al. The target implosion should be robust against the implosion non-uniformities for the stable reactor operation.

In the HIBs energy deposition region in a DT fuel target a wide density valley appears, and in the density valley a part of the HIBs deposited energy is converted to the radiation and the radiation is confined in the density valley. The converted and confined radiation energy is not negligible, and it would be the order of ~100 kJ in a HIF reactor-size DT target. The confined radiation in the density valley contributes also to reduce the non-uniformity of the HIBs energy deposition [12].

The HIB must be transported in a fusion reactor, which would be filled by a debris gas plasma. The reactor radius would be 3~5 m or so. From the beam exit of the accelerator the HIB should be transported stably. The HIB ion is rather heavy. For example, $Pb^+$ ion beams could be a promising candidate for the HIF driver beam. Fortunately, the heavy ions are transported almost ballistically with straight trajectories in a long distance. Between the HIB ions and the background electrons, beam-plasma interactions occur: the two-stream and filamentation instabilities may appear, and simple analyses confirm that the HIBs are almost safe from the instabilities' influences [20].

The HIB uniform illumination study presented that the target implosion uniformity requirement requests the minimum HIB number: details HIBs energy deposition on a direct-drive DT fuel target shows that the minimum HIBs number would be the 32 beams [21]. The 3-dimensional detail HIBs illumination on a HIF DT target is computed by a computer code of OK [22]. The HIBs illumination non-uniformity is also studied in detail. One of the study results shows that a target misalignment of ~100μm is tolerable in fusion reactor to release the HIF energy stably. The target alignment in a fusion reactor was also studied [23]. The research results demonstrate that a target alignment error can be controlled by a magnetic system at the target injection gun [23].

In the perspective article the merits of the HIF energy system are focused toward our bright human life.

**2. Issues in HIF**

In this section key issues in HIF are summarized first [24]. The fuel target design should be conducted further toward a robust fuel implosion, ignition and burning. The HIF target design is quite different from the laser fusion target due to the relatively long range (the order of ~0.5-1mm) of the energy deposition [1, 21, 22]. An example direct-drive fuel target has a thicker energy absorber, corresponding to the ion stopping range of ~0.5-1mm. The target should be compressed to about one thousand of the solid density to reduce the driver energy and to enhance the fusion reactions [25]. The target should be robust against the small non-uniformities caused by the driver beams' illumination non-uniformity [18, 19, 22], a fuel target alignment error in a fusion reactor [23], the target fabrication defect, et al. The ICF reactor operation frequency would be 10~15 or so.

The HIB stopping range is rather long, and the HIB beam energy is deposited mainly at the end area of the beam ion stopping range due to the Bragg peak effect, which is originated from the nature of the Coulomb collision. The interaction of the HIB ions could be utilized to enhance the HIB preferable characteristics. The HIB ion interaction is relatively simple, and is almost the classical Coulomb collision, except the plasma range-shortening effect [6, 7, 26]. Therefore, the HIB energy deposition profile is well defined in the HIF target. A typical HIB ion species would be $Pb^+$ or $Cs^+$ or so. For $Pb^+$ ions the appropriate Pb ion energy would be about 8GeV or so.

The HIBs illumination scheme was studied intensively to realize a uniform energy deposition in a HIF target. The ICF target implosion uniformity must be less than a few % [18, 19]. The uniformity requirement must be fulfilled to release the fusion energy. The multiple HIBs should illuminate the HIF target with a highly uniform scheme during the imploding DT shell acceleration phase.

In addition, the HIB pulse shape should be also designed to obtain a high implosion efficiency $\eta_{imp}$. The simple and typical pulse shape consists of a low-intensity foot pulse, a ramping part to the peak intensity and the high-intensity main pulse [15]. The foot pulse generates a weak shock wave in the target material and the DT fuel, and the first shock wave kicks the low-temperature DT liquid fuel inward. When no foot pulse is used, the main pulse with the high intensity generates a strong first shock wave inside of the DT fuel and increases the DT adiabat. The DT fuel preheat would be induced and the efficient fuel compression would be realized. The foot pulse length and the ramping time are designed to reduce the entropy increase. The first weak shock wave is not caught by the second and third stronger shocks inside of the DT fuel layer. At the inner edge of the DT fuel layer the shocks should be overlapped so that the efficient fuel acceleration and compression are accomplished during the implosion. A stronger shock wave increases the entropy more from the fluid ordered motion than the weak shock wave. The detail pulse shape should be designed for each target design.

The DT fuel stable compression is another key issue in ICF. In the recent NIF target experiments in Livermore, CA, U. S. A., they have succeeded to compress the DT fuel to a few thousands of the solid density in ICF [27, 28]. The highly compressed DT fuel met a DT fuel mix, that would be induced by the R-T instability. The R-T instability and the fuel mix are caused by the implosion non-uniformity, the driver laser illumination non-uniformity, etc. The DT fuel implosion stability should be studied carefully in ICF. The DT fuel compression dynamics itself in HIF is the same as that in the laser fusion, though the driver target interaction is quite different from that in the laser fusion.

Each target should be injected into the reactor, and must be situated at the reactor center. The target alignment or positioning error in the reactor should be minimized to reduce the HIBs illumination non-uniformity. The experimental results show that the target alignment error would be minimized to ~100μm or so [23]. This issue is also studied for a real reactor design. On the other hand, the target should be designed to be robust against the target alignment of ~ 100μm.

The HIF reactor would have a dilute reactor gas inside the reactor chamber after each fuel target shot. The reactor operation frequency would be 10~15 Hz in HIF. In HIF the debris gas would supply cold electrons to compensate the HIB space charge in the vicinity of the fuel target and / or during the HIBs transport in the reactor. The reactor radius may be 3-5m or so. The debris gas dynamics should be also studied [23, 24].

The HIB transport in the reactor is another key issue. The HIB ion mass is large so that the ion trajectory is almost straight between the HIB accelerator exit to the reactor center. However, each HIB carries a large current of ~1kA or so. The remaining large current and its self-charge would provide a slight HIB radial expansion [20]. The final neutralized HIB transport should be also studied.

The reactor design is also another key issue in ICF [2-5]. The first wall could be a wet wall with a molten salt or so or a dry wall. The reactor design must accommodate a large number of HIBs beam port, for example, 32 beam ports. At the first wall and the outer reactor vessel the beam port holes should have mechanical shutters or so to prevent the fusion debris exhaust gas toward the accelerator upstream. In addition, the target debris remains inside of the first wall or mixes with the liquid molten salt, which may be circulated. The target debris treatment should be also studied as a part of the reactor design. The tritium (T) also remains inside of each target after its burning. Usually about 30% or so of the DT fuel is reacted and depleted in the target during the burning process. So a large part of T in each target is mixed in the reactor gas and would be melted in the liquid first wall. The rest T and the radio-activated target materials must be distributed inside of the reactor vessel right after the target burning, and they would not accrete in a large lump in the reactor. The distributed radioactive materials must be collected and separated in the fusion reactor system safely.

The HIB accelerators have the high controllability and flexibility for the particle energy, the beam pulse shape, the pulse length, the ion species, the beam current, the beam radius, the beam focusing and the beam axis motion. The accelerators are operated daily with a high frequency stably. This is a very preferable feature for the fusion reactor system [29]. However, the high-current operation may need additional studies for the fusion reactor design. The ~kA HIB may induce an additional HIB divergence, a beam loss and an electron cloud generation in the accelerator. The high-current and high-charge HIB generation and transport should be studied carefully to avoid the uncertainty in the HIB accelerator. The high-current HIB accelerator type should be also another issue in HIF accelerators [30, 31]. In the accelerator system, each beam is generated first from the ion source, and near the final beam transport to the HIF reactor each HIB may be compressed longitudinally. The HIB pulse shape in Fig. 2 may impose a requirement of the HIB longitudinal length. For example, an 8GeV Pb ion has the speed of ~0.29c, where c is the speed of light. The 20ns HIB pulse length is about 1.7m. At the last stage of the accelerator, the HIB may be compressed or bunched in the longitudinal direction. The HIB longitudinal bunching may be also required.

In addition to single heavy ions as the HIF driver, we presently witness a reappraisal of interest for the cluster ion beam (CIB) approach to heavy ion driven ICF, including cluster ion acceleration, cluster fragmentation and target correlated stopping of cluster ions. CIB driven HIF displays a considerable potential for achieving Directly as well as indirectly driven ICF through vastly enhanced and correlated target ion stopping altogether with a lower accelerating tune shift [32, 33].
We intend to stress a recent proposal switching from C[60, n+] linearly accelerated ions to circularly accelerated Si[108, 8+]ions through a suitable combination of induction synchrotron technology with an extension of the relativistic electron Microtron to GeV energies for very heavy particles [34]. This program is currently developed on the KEK facility (Tsukuba, Japan). Focusing on the strongly correlated stopping featured by the charged debris resulting from the impacting cluster ion projectiles hurling to the ICF pellet at ~10 MeV/amu, one thus expects that the direct drive approach could be managed with a higher and efficient hydro compression, while the indirect one could operate on a higher adiabat with a 500-600 eV radiative temperature. The promising prospects also rely on significant advances achieved on the cluster ion sources design as well as on efficient laser conditioning of the Coulomb explosion of the cluster ion projectiles in a target [35]. The CIB merit would be found in the lower $q/A$, which may be relaxing the accelerator design. The CIF ion source and the stable cluster acceleration should be studied further toward the CIF realization based on experiments and theories [36].

In the following Sections, we pick up and discuss key factors on accelerators, laser ion source for accelerators and reactor designs in HIF.

3. Accelerator for HIF

As shown in Sec. 1, the concept of particle beam driven inertial fusion was proposed in 1970's. Almost 50 years have passed since early discussions on the compatibility of particle beam drivers for the fusion reactor system. During the period, the driver particles shifted gradually from light particles to heavier ions [37]. Although beam power is the product of beam energy and the current, the beam particle energy is restricted to an upper limit due to an appropriate range in the outer region of the fusion fuel target. When light ions such as proton or lithium ions (LIBs: Light Ion Beams) are used as the driver, the upper limit of the beam energy is a 10MeV. In the case of LIBs, the beam current of 10MA level is required to drive the fuel with a beam power of 100TW level. The LIBs require an extremely high current beam source driven by a low impedance pulse power device, and the huge current must be extracted from a single gap (a pulse powered diode). Hard works and efforts had been made in the world to improve the controllability of high current LIBs. However, pulse power diodes were difficult to control and could not make repetitive operation due to a direct connection between the single acceleration gap and the low impedance power source.

From a point view of beam controllability during the extraction and propagation, the beam current should be as low as possible. Namely, to reduce the current level and increase the controllability, the driver beam should be as heavy as possible. The particle beam fusion shifted from LIBs to schemes using HIBs. The research efforts have shifted to issues concerning high power HIB accelerators.

The factors of drivers, which should be considered for the fusion reactor, are the construction cost, the repetition capability, and the controllability. The cost and scale of accelerator steadily increase with heavier projectiles. For a fusion driver, the total energy of the ions must be a MJ level. That means that a HIB with a 10GeV particle energy has the total number of ~$10^{15}$ level of ions. The ions should be repetitively extracted, accelerated, manipulated, transported, and irradiated to the target surface with the time scale of less than a 10nsec.

The figure of merits of the heavier ions is reducing the space charge effects. However, intense HIB accelerators were extremely expensive. Particularly to prepare an additional induction linac for the final bunch compression brings extraordinal increase in its cost and size. At the present, we have still no concrete final solution for appropriate accelerator scheme for the fusion reactor.

Recently a novel accelerator scheme based on a concept of induction synchrotron has been proposed [38], which increases remarkably the flexibility of beam handling due to separation of beam acceleration and confinement in the moving direction. The new concept extends the range of possible projectile ions for the acceleration up to giant clusters. It reduces the space charge issues of the beam due to the system flexibility and increase of the particle energy. A scheme was proposed to accelerate giant cluster ions toward 120GeV based on a two-way multiplex configuration [38]. The accelerator is composed of five stories, two-way beam lines and a permanent stacking ring. In the two-way accelerator ring, 5 forward and 5 backward super bunches share the induction cells for acceleration and confinement, in which the forward bunches are accelerated with a set voltage and the backward super bunches with a reset voltage of the repetitive induction modulators.

The beam bunches are accelerated and modulated in the ring with multiple beam guiding beam lines with ten-fold symmetry. As a result of the manipulation and drift compression in the ring, the long bunch rotates in the longitudinal phase space and makes a 100s of short multiple bunches, which delivers the beam power on a fuel target, placed at the ring center with a pulse length of 10nsec.

The two-way, multiplex, and the ring with the symmetrical guiding lines configuration is crucial to reduce the machine size and cost. That is, 10 accelerator rings are accommodated in a single ring. Until recently, indirect drive of the target is considered to be the only scheme of HIB fusion system. However, the compact and economical multiplex accelerator system together with the new irradiation scheme [24] should enable us to reopen the exploration of the direct drive scheme by the multiple beam illumination. In other words, the cost-effective accelerator scheme has a huge potential for both indirect and direct irradiation schemes of particle beam fusion.

An advanced conceptual design study on the fusion reactor scheme using the two-way multiplex accelerator is under development and the details will be clarified in the near future [39]. The points that we have to take care about, for optimizing the ion mass of the driver, are a beam loss during the acceleration of the beam particle to the appropriate energy, and a rational-symmetry target design compatible to the irradiation scheme.

4. Laser ion source for the particle accelerator

The HIF scenarios proposed require low charge state ion beams of ~ampere class of heavy species. The laser ion source (LIS) is recognized as one of the promising candidates for ion beam providers, since it can deliver a high brightness heavy ion beam to accelerator. The design of LIS for HIF depends on the accelerator structure and accelerator complex following the source [40]. The first descriptions of laser ion sources can be found in the articles from the 1960's [41, 42]. The laser was invented in 1960, and the first laser ion source was proposed within a decade. Therefore, the laser ion source was one of the earliest applications of lasers. An intense pulse of laser light is focused on a solid target placed inside a vacuum vessel. The laser energy is used to ionize the target material, and an ablation plasma is induced. Since the laser irradiation period is very short, typically less than a few tens of nanoseconds, the heated plasma does not have time to expand during the pulse, and its density remains high. No extra confinement forces, such as magnetic field, are needed. After the laser irradiation, the plasma expands slowly and at the same time leaves from the target surface. Therefore, the gravity center of the expanding plasma has a velocity perpendicular to the target surface. When the head of the plasma plume reaches the extraction voltage gap, the ion beam formation begins. This process continues until the end of the plasma plume reaches the extraction electrode. Although the laser irradiation is very short (~10 ns), the pulse width of the extracted ion beams can be extended to the microsecond scale.

For the ion species of the driver beam, we propose to use single-isotopic elements, such as $^{89}$Y, $^{93}$Nb, $^{103}$Rh, $^{127}$I, $^{133}$Cs, $^{197}$Au and $^{209}$Bi. For example, if we use Pb, which is a typical heavy mass element, it has four isotopes, where $^{208}$Pb occupies only 58% of its natural abundance. The other three isotopes have light masses and may cause beam losses somewhere in the accelerator chain. The chemical stability of the species, used as a laser target must be also considered. For example, Ti and Zr trap a large amount of ambient gas and these impurities may create unwanted ions. Therefore, it is recommended to use Au$^+$ ions for the HIF driver beams. If a higher charge-to-mass ratio ion is preferred in order to shorten the accelerator, we suggest Nb$^+$ rather than Au$^{2+}$. Because the laser irradiation conditions to produce charge state 2+ usually produce ions in adjacent charge states. State of the art techniques delivers 100 mA class Au$^+$ or 200 mA class Nb$^{2+}$ beams to induction or RF linear accelerators. To maximize the efficiency of the entire accelerator system, the beam current profile needs to be tailored, and more advanced beam shaping systems are under investigations. Recently, the beam stability analysis of the plasma transport inside the solenoidal field was performed. When using the solenoid in a LIS, the beam current density can be increased. The beam current can be also adjusted stably by changing the strength of the solenoid magnetic field within a proper range [43].

5. Summary and HIF energy perspective

As we have discussed above, the ion-beam based DT inertial fusion has preferable features to release the fusion energy to sustain our human society with respect to the repetitive driver operation, the driver high efficiency, the 100% HIB energy deposition in the fuel target, the concrete HIB ion energy deposition profile in the fuel target deep layer, etc. The crucial point in ICF was the fuel implosion and ignition of the DT shell target. However, the recent NIF experiments have ensured that the DT shell target is compressed to a high density more than 1000 times of the solid density [27, 28]. The crucial point of the high-density compression was demonstrated by the NIF results. In HIF 4-5MJ of the total HIB driver energy would be needed to obtain a sufficient fusion energy output stably and repetitively.

As discussed above, HIF has no fatal problems to construct a HIF reactor system except technical issues relating to the T treatment, the treatment of the radioactive materials, the costs, the electric power flow of the HIF reactor plant system, etc. In the near future we need to have an advanced conceptual design including the new progresses in the HIF researches as well as other remaining issues.

The intense ion beam provides also a unique tool to investigate unique physics in dynamic instability stabilization and in High Energy Density Physics, and provides a promising driver to study science in ion beam inertial fusion energy as our reliable future energy source, which is discussed and summarized in this paper. Beam ions deposit their energy inside a material, and so a warm or a high-energy area is created inside a material. One good example of the ion beam applications is an ion beam cancer therapy, in which a cancer inside a human body is directly illuminated without a serious damage of normal cells. The ion beam also has another unique feature of a precise controllability of the particle energy, the beam direction, the beam pulse duration, the beam pulse shape and also a wobbling capability by the ion beam axis precise rotation or oscillation. The wobbling capability of the ion beam presents another innovative tool to stabilize instabilities dynamically in plasmas and fluids [12-15]. For example, the Rayleigh-Taylor instability growth would be controlled and stabilized significantly by the ion beam wobbling.

At present an experimental device of NDCX II at Berkeley, CA, U. S. A. works on the HIB accelerator physics study [44]. The FAIR (Facility for Antiprotons and Ion Research) project has started at Darmstadt, Germany [45]. FAIR is oriented to basic physics to understand the structure of matter, the evolution of the universe, etc. including the plasma physics. The HEDP (High Energy Density Physics) studies would be included in the plasma physics project in FAIR. The HIAF (High Intensity heavy ion Accelerator Facility) project in China has been planned for HIF and HEDP studies [46]. The HIAF construction has started in 2018 in Huizhou City. The maximum energy and power in one beam pulse could reach 100 kJ and 1 TW, respectively. With such the high power heavy ion beam, a large volume of matter with a temperature over 10eV could be generated by a direct heating of a solid target. In addition, in the current design HIAF contains a colliding terminal with two high-power-beams at each side, which would provide a promising opportunity for HIF studies [47-49]. In addition to the large accelerator facilities shown above, theoretical, simulation and small-scale experimental studies have been performed in various places. The researchers on HIF have been conducted a large number of research labs. and researchers in U. S. A., Japan, Germany, France, Russia, China, Italy, Spain, Israel, Kazakhstan, etc. [50-53].

**Acknowledgments**

The work was partly supported by JSPS, MEXT, CORE (Center for Optical Research and Education, Utsunomiya University), ILE / Osaka University, and CDI (Creative Department for Innovation, Utsunomiya University). The authors also would like to extend their acknowledgements to friends in HIF research group in Japan, in Tokyo Inst. of Tech., Nagaoka Univ. of Tech., KEK and also in HIF-VNL, U.S.A. and in Utsunomiya Univ. Our former and present graduate students have contributed partly to the HIF studies, and they are also acknowledged. The authors would also like to express their appreciations to Prof. Colin Danson and Prof. Jianqiang Zhu, the Co-Editors-in-Chief of HPLSE, who encouraged us to prepare the paper.

**References**

1. M. J. Clauser, Phys. Rev. Lett. **35**, 848 (1975).
2. D. Böhne, I. Hofmann, G. Kessler, G.L. Kulcinski, J. Meyer-ter-Vehn, U. von Möllendorff, G.A. Moses, R.W. Müller, et al., **73**, 195 (1982).
3. T. Yamaki, et al., Nagoya University, Report **IPPJ-663**, (1985).
4. I. Hofmann, et al., GSI Report **GSI-98- 06**, (1998).
5. K. Niu and S. Kawata, Fusion Technology **11**, 365 (1987).
6. J. F. Ziegler, J. P. Biersack, and U. Littmark, "The Stopping and Range of Ions in Matter",

volume **1**, (Pergamon, New York, 1985).

7. T. A. Mehlhorn, J. Appl. Phys. 52, 6522 (1981).

8. D. A. Callahan-Miller and M. Tabak, Nuclear Fusion **39**, 883 (1999).

9. R. C. Arnold, et al., Nucl. Inst. Meth. **199**, 557 (1982).

10. A. R. Piriz, A. R., N. A. Tahir, D. H. H. Hoffmann, and M. Temporal, Phys. Rev. E **67**, 017501 (2003).

11. H. Qin, R. C. Davidson, and B. G. Logan, Phys. Rev. Lett. **104**, 254801 (2010).

12. S. Kawata, T. Sato, et al., Laser Part. Beams **11**, 757 (1993).

13. S. Kawata, Phys. Plasmas, **19**, 024503 (2012).

14. S. Kawata and T. Karino, Phys. Plasmas, **22**, 042106 (2015).

15. R. Sato, S. Kawata, T. Karino, et al., Sci Rep **9**, 6659 (2019). https://doi.org/10.1038/s41598-019-43221-7

16. S. E. Bodner, Phys. Rev. Lett. **33**, 761 (1974).

17. H. Takabe, K. Mima, et al., Phys. Fluids **28**, 3676 (1985).

18. M. H. Emery, et al., Phys. Rev. Lett. **48**, 253 (1982).

19. S. Kawata, and K. Niu, J. Phys. Soc. Jpn. **53**, 3416 (1984).

20. S. Kawata, et al, , Nucl. Inst. Meth. Phys. Res. A **544**, 98 (2005).

21. S. Miyazawa, A. I. Ogoyski, et al., Phys. Plasmas, **12**, 122702-1-9 (2005).

22. A. I. Ogoyski, T. Someya and S. Kawata, Computer Phys. Commun. **157**, 160 (2004); A. I. Ogoyski, et al., Computer Phys. Commun. **161**, 143 (2004); A. I. Ogoyski, et al., Computer Phys. Commun. **181**, 1332 (2010).

23. T. Kubo, T. Karino, H. Kato and S. Kawata, IEEE Trans. Plasma Sci., **47**, 2 (2019); H. Nakamura, T. Kubo, T. Karino, H. Kato, S. Kawata, High Energy Density Phys. **35**, 100741 (2020).

24. S. Kawata, T. Karino, and A. I. Ogoyski, Matter and Radiation at Extremes **1**, 89 (2016); https://doi.org/10.1016/j.mre.2016.03.003

25. S. Atzeni and J. Meyer-ter-Vehn, "*The Physics of Inertial Fusion"* (International Series

of Monographs on Physics, 2009).

26. S. Ichimaru, "*Statistical Plasma Physics*", (Westview Press, 2004).

27. O. A. Hurricane, D. A. Callahan et al., Nature **506**, 343 (2014).

28. H.-S. Park, O. A. Hurricane, D. A. Callahan, et al., Phys. Rev. Lett. **112**, 055001 (2014).

29. R. O. Bangerter, Nucl. Instr. Meth. A **415**, 3 (1998).

30. M. Okamura et al., Nucl. Instr. Meth. A **606**, 94 (2009).

31. K. Takayama, R. J. Briggs, et al., "*Induction Accelerators*", (Springer-Verlag Berlin Heidelberg, 2011).

32. C. Deutsch, A.Bret, S.Eliezer, J.M.Martinez-Val and N.A Tahir, Fusion Technology **31**,1(1997)

33. N. A. Tahir, K. J. Lutz, O. Geb, J. A. Maruhn, C. Deutsch and D. H. H. Hoffmann, Phys. Plasmas **4**,706 (1997).

34. K. Takayama and K. Horioka, Proceedings of HIF2016, (2016).

35. G. Q. Wang, et al, Phys. Rev. A **86**, 043201(2012).

36. S. Kawata, C. Deutsch, and Y. J. Gu, Phys. Rev. E **99**, 011201(R) (2019).

37. K. Horioka, Matter and Radiation at Extremes **3**, 12 (2018)

38. K. Takayama, and K. Horioka, Proceedings of HIF 2016.

39. K. Takayama et.al., in preparation.

40. M. Okamura, Matter and Radiation at Extreme **3**, 61 (2018).

41. N. J. Peacock, and R. S. Pease, Br. J. Appl. Phys. 2, 1705 (1969).

42. Y. A. Byckovsky, V. F. Eliseev, Y. P. Kozyrev, and S. M. Silnov, Sov. Patent **324**, 938 (1969).

43. T. Karino, M. Okamura, et al., Rev. Sci. Instrum. **91**, 053303 (2020). doi: 10.1063/1.5128512

44. J. J. Barnard, et al., Nuclear Instruments Methods Phys. Res. A **733**, 450 (2014).

45. FAIR, http://www.fair-center.eu/public/what-is-fair.html

46. J. C. Yang, J. W. Xia, G. Q. Xiao, H. S. Xu, H. W. Zhao, et al., Nucl. Instr. Meth. B **317**,

263 (2013).

47. Y. Zhao, et al, submitted to Scientia Sinica in Physica, Mechanica & Astronomica, (2020).

48. J. Ren, Y. Zhao R. Chen, et al., Nucl. Instr. Meth. B **406**, 703 (2017).

49. R. Cheng, et al, Matter and Radiation at Extremes, **3**(2), 85 (2018).

50. D. Wu, W. Yu, Y-T. Zhao. et al., Phys. Rev. E **100**(1), 013208 (2019).

51. L. Zhang,Y. Zhao, J. Ren, et al., IEEE Trans. Plasma Sci. **47**, 853 (2019).

52. L. Zhang, Y. T. Zhao, J. R. Ren, et al., Phys. Plasmas, **25**, 113108 (2018).

53. B. Z. Chen, D. Wu, J. R. Ren, et al., Phys. Rev. E, accepted (2020).